\documentclass{article}
\usepackage{spconf,amsmath,graphicx,cite}
\usepackage{enumitem}
\usepackage{threeparttable}
\usepackage{multirow}
\usepackage{multicol}
\usepackage{booktabs}
\usepackage{algorithm}
\usepackage{algorithmic}
\usepackage[dvipsnames]{xcolor}
\usepackage{enumerate}
\usepackage{diagbox}
\usepackage{comment}
\usepackage{lipsum}
\usepackage[utf8]{inputenc}
\usepackage{color}
\usepackage{amssymb}
\usepackage{bm}
\usepackage{caption}
\usepackage{subcaption}

\PassOptionsToPackage{hyphens}{url}\usepackage{hyperref}
\hypersetup{colorlinks=true}

\newcommand{\pluseq}{\mathrel{+}=}
\newcolumntype{C}{>{\centering\arraybackslash}m{6.5ex}}
\newcolumntype{D}{>{\centering\arraybackslash}m{7.5ex}}

\title{Factorized Blank Thresholding for Improved Runtime Efficiency of Neural Transducers}
\name{Duc Le, Frank Seide, Yuhao Wang, Yang Li, Kjell Schubert, Ozlem Kalinli, Michael L. Seltzer}
\address{Meta, USA\\ \small{\texttt{duchoangle@meta.com}}}
\begin{document}
\ninept
\maketitle
\begin{abstract}
We show how factoring the RNN-T's output distribution can significantly reduce the computation cost and power consumption for on-device ASR inference with no loss in accuracy. With the rise in popularity of neural-transducer type models like the RNN-T for on-device ASR, optimizing RNN-T's runtime efficiency is of great interest. While previous work has primarily focused on the optimization of RNN-T's acoustic encoder and predictor, this paper focuses the attention on the joiner. We show that despite being only a small part of RNN-T, the joiner has a large impact on the overall model's runtime efficiency. We propose to utilize HAT-style joiner factorization for the purpose of skipping the more expensive non-blank computation when the blank probability exceeds a certain threshold. Since the blank probability can be computed very efficiently and the RNN-T output is dominated by blanks, our proposed method leads to a 26-30\% decoding speed-up and 43-53\% reduction in on-device power consumption, all the while incurring no accuracy degradation and being relatively simple to implement.
\end{abstract}
\section{Introduction}
\label{sec:intro}

The neural transducer, specifically the recurrent neural network transducer (RNN-T)\footnote{Despite the name, RNN-T does not necessarily utilize recurrent units.}\cite{Graves12}, has become increasingly popular for automatic speech recognition (ASR)~\cite{GravesMohamedHinton13, RaoSakPrabhavalkar17,YehMahadeokarKalgaonkar19, ZhangLuSakEtAl20, ShiWangWuEtAl21}. RNN-T is especially attractive for on-device applications due to its compact sizes, all-neural architecture, and ease of streaming, which enable relatively accurate transcription with low latency and small memory/computational footprint~\cite{Prabhavalkar17,HeSainathPrabhavalkarEtAl19,shangguan2019optimizing}. Given the limited on-device computation budget, optimizing RNN-T's runtime efficiency is of considerable interest.

Much of the previous work on this topic has focused on optimizing RNN-T's acoustic encoder (analogous to the acoustic model in a traditional hybrid ASR system), which is usually the largest part of the model. Example optimizations include adopting a more efficient autoregressive unit~\cite{shangguan2019optimizing} or shifting to non-autoregressive models (attention-based and/or convolution-based) to better take advantage of batched processing across multiple time steps~\cite{Gulati2020conformer,ShiWangWuEtAl21,shi2022streaming}. The prediction network or \emph{predictor} (analogous to the language model) has also received several optimizations, such as moving from autoregressive to non-autoregressive (stateless) architectures~\cite{Ghodsi20stateless,Rohit21LessIsMore} or simple embedding lookup tables~\cite{Botros21TiedReduced}.

RNN-T's joint network or \emph{joiner} (analogous to the decoder), on the other hand, has not been explored much from a runtime efficiency perspective. Typically consisting of just one fully connected (FC) layer, the joiner is computationally cheap compared to the encoder and predictor. Despite of that, however, the joiner has a surprisingly large influence on the on-device decoding speed as well as power consumption, as it is executed very frequently during decoding. 

In Hybrid Autoregressive Transducer (HAT) \cite{VarianiRybachAllauzenEtAl20}, the authors proposed to factorize the joiner into separate blank and non-blank portions that are computed separately, aiming to estimate internal language model (LM) scores for more accurate integration of external LMs~\cite{VarianiRybachAllauzenEtAl20,Sainath21Efficient,Chen22factorized}. We propose to use the same factorization, but for the purpose of improving runtime efficiency.

In this paper, we leverage the joiner-factorization idea and the ``spiky" nature of RNN-T's output as follows. We first compute only the blank probability separately; if the remaining non-blank probability mass is below a threshold, skip the significantly more expensive non-blank computation altogether. We show that this simple method can avoid roughly \textbf{two-thirds} of non-blank joiner invocations, resulting in \textbf{26-30\%} improvement in decoding speed and \textbf{43-53\%} reduction in on-device power consumption due to reduced memory bandwidth. Crucially, our method incurs no Word Error Rate (WER) degradation, neither on Librispeech nor on large-scale in-house data. To the best of our knowledge, this is the first work to explore optimizing RNN-T's joiner to improve runtime efficiency.

\section{Factorized Blank Thresholding}
\label{sec:method}

\subsection{Factorized Joiner}
\label{ssec:model_arch}

\begin{figure}[t]
  \centering
  \includegraphics[width=\columnwidth]{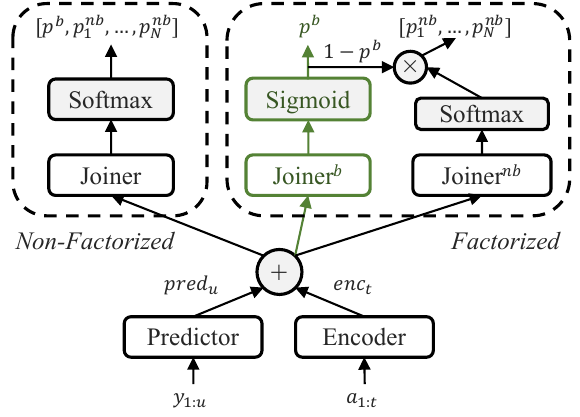}
  \caption{RNN-T with non-factorized vs. HAT-style factorized joiner.}
  \label{fig:model}
\end{figure}

\begin{algorithm}[t] % enter the algorithm environment
\caption{RNN-T Beam Search \textcolor{red}{with Blank Thresholding}} % give the algorithm a caption
\label{rnnt_beamsearch_algo} % and a label for \ref{} commands later in the document
% Proposed modifications in Algorithm 1 from \cite{rnnt-graves} are in \textcolor{red}{red}.
% Modifications are in \textcolor{red}{red}.
\begin{algorithmic} % enter the algorithmic environment
\STATE \textbf{Initialize:} \hspace{0.05in} $B$ = $\{\varnothing\}$; \hspace{0.05in} $\mathrm{\mathrm{Pr}}(\varnothing)$ = 1
\FOR {$t$ = 1 to $T$}
    \STATE $A = B$
    \STATE $B = \{\}$
    \FOR {$\bm{y}$ in $A$}
    \STATE $\mathrm{\mathrm{Pr}}(\bm{y}) \pluseq \sum_{\hat{\bm{y}}\in \mathrm{pref}(\bm{y})\cap A} \mathrm{\mathrm{Pr}}(\hat{\bm{y}}) \mathrm{\mathrm{Pr}}(\bm{y}|\hat{\bm{y}},t) $
    \ENDFOR
    \WHILE{$B$ contains less than W elements more
probable than the most probable in $A$}
        \STATE $\bm{y}^\ast = $ most probable in $A$ 
        \STATE Remove $\bm{y}^\ast$ from $A$
        \STATE \textcolor{red}{Compute factorized blank probability $\mathrm{Pr}(\varnothing|\bm{y}^\ast,t)$}
        \STATE $\mathrm{Pr}(\bm{y}^\ast) = \mathrm{Pr}(\bm{y}^\ast) \mathrm{Pr}(\varnothing|\bm{y}^\ast,t)$
        \STATE Add $\bm{y}^\ast$ to $B$
        \textcolor{red}{\IF{$\mathrm{Pr}(\varnothing|\bm{y}^\ast,t) \le p^{\text{threshold}}$}
            \STATE Compute factorized non-blank probabilities $\mathrm{Pr}(k|\bm{y}^\ast,t)$
            \textcolor{black}{\FOR{$k \in Y$}
                \STATE $\mathrm{Pr}(\bm{y}^\ast\hspace{-0.3em}+\hspace{-0.3em}k) = \mathrm{Pr}(\bm{y}^\ast) \textcolor{red}{(1\hspace{-0.3em}-\hspace{-0.3em}\mathrm{Pr}(\varnothing|\bm{y}^\ast,t))} \mathrm{Pr}(k|\bm{y}^\ast,t)$
                \STATE Add $\bm{y}^\ast + k$ to $A$
            \ENDFOR}
        \ENDIF}
    \ENDWHILE
    \STATE Remove all but the $W$ most probable from $B$
\ENDFOR
\RETURN  $\bm{y}$ with highest log $\mathrm{Pr}(\bm{y})/|\bm{y}| $ in $B$ 
\end{algorithmic}
\label{alg:beam_search}
\end{algorithm}

In a standard RNN-T with non-factorized joiner, the posterior probability $\mathrm{Pr}(\bar{y}|\bm{y}_{1:u},t)$, where $\bar{y}$ is either the next non-blank output token $y_{u+1}$ or the blank token $\varnothing$, $\bm{a}_{1:t}$ is the sequence of observed audio frames, and $\bm{y}_{1:u}$ is the sequence of previously emitted non-blank tokens, is computed as follows:
\begin{align}
    &\mathrm{enc}_t = \texttt{Encoder}(\bm{a}_{1:t})\\
    &\mathrm{pred}_u = \texttt{Predictor}(\bm{y}_{1:u})\\
    &\mathrm{join}_{t,u} = \texttt{Joiner}(\mathrm{enc}_t + \mathrm{pred}_u) \in {\rm I\!R}^{N+1}\\
    &\mathrm{Pr}(\bar{y}|\bm{y}_{1:u},t) = \texttt{Softmax}(\mathrm{join}_{t,u})_{\bar{y}}
    %&\mathrm{Pr}(\bar{y} = i|\bm{y}_{1:u},t) = \texttt{Softmax}(\mathrm{join}_{t,u})_i
\end{align}
\noindent In this formulation, the blank and non-blank probabilities are computed together using a single network $\texttt{Joiner}$, which has an output dimension of $N+1$, where $N$ is the number of non-blank tokens.

In an RNN-T with HAT-style factorized joiner~\cite{VarianiRybachAllauzenEtAl20}, the posterior probability is computed as follows:
\begin{align}
    &\mathrm{join}^\mathrm{b}_{t,u} = \texttt{Joiner}^b(\mathrm{enc}_t + \mathrm{pred}_u) \in {\rm I\!R}\\
    &\mathrm{join}^\mathrm{nb}_{t,u} = \texttt{Joiner}^{nb}(\mathrm{enc}_t + \mathrm{pred}_u) \in {\rm I\!R}^N\\
    &p^\mathrm{b} = \texttt{Sigmoid}(\mathrm{join}^\mathrm{b}_{t,u})\\
    &\bm{p}^\mathrm{nb} = (1 - p^\mathrm{b}) \cdot \texttt{Softmax}(\mathrm{join}^\mathrm{nb}_{t,u})\\
    &\mathrm{Pr}(\bar{y}|\bm{y}_{1:u},t) = [p^\mathrm{b};\bm{p}^\mathrm{nb}]_{\bar{y}}
    %&\mathrm{Pr}(\bar{y}=i|\bm{y}_{1:u},t) = [p^\mathrm{b};\bm{p}^\mathrm{nb}]_i
\end{align}
where $\texttt{Joiner}^b$ is the blank joiner, $\texttt{Joiner}^{nb}$ is the non-blank joiner, $p^\mathrm{b}$ is the blank posterior probability, and $\bm{p}^\mathrm{nb}$ is the non-blank posterior probability vector. Here, the blank and non-blank probabilities are computed separately using two different joiners. Crucially, the blank probability can be computed much more efficiently since $\texttt{Joiner}^b$ has an output dimension of 1 (the output projection reduces to a dot product), while $\texttt{Joiner}^{nb}$ has an output dimension of $N$ (several hundred). Figure~\ref{fig:model} illustrates the differences between the non-factorized and HAT-style factorized joiners.

\subsection{Beam Search Decoding with Blank Thresholding}
\label{ssec:beam_search}

Similar to acoustic models trained with the Connectionist Temporal Classification (CTC) loss~\cite{GravesFernandezGomezEtAl06}, RNN-T is known to produce ``spiky" outputs: $\max_{\bar{y}} \mathrm{Pr}(\bar{y}|\bm{y}_{1:u},t) \approx 1$, with blank dominating in most time steps. Thus, if $p^\mathrm{b} \approx 1$, we assume the $\bm{p}^\mathrm{nb}$ are all zero, and can skip computing them altogether. Algorithm~\ref{alg:beam_search} shows a modified version of Graves' original RNN-T beam search~\cite{Graves12} with blank thresholding via the factorized joiner. Our modifications are in {\color{red}red}.

As can be seen, our modified beam search only performs non-blank computation and extension if the blank probability falls below the threshold $p^{\text{threshold}}$.
As $p^{\text{threshold}}$ gets smaller, more non-blank computations will be skipped. In practice, we apply this thresholding logic in conjunction with the extra beams described in~\cite{Jain19decoding} to further improve decoding speed.

The compute savings from blank thresholding come from skipping much of the joiner computation. Thus, while it is possible to use blank thresholding with a standard non-factorized joiner to speed up beam search, this has a negligible impact in practice.

\section{Experimental Setup}
\label{sec:exp}

\subsection{Datasets and Metrics}
\label{ssec:data}

\subsubsection{Librispeech}
\label{sssec:librispeech}

\begin{table}[t]
\centering
\begin{tabular}{ | c | C | D | C | D | }
    \hline
    \multirow{2}{*}{\textbf{Component}} & \multicolumn{4}{c|}{\textbf{Model}} \\
    \cline{2-5}
    & \textbf{\emph{NF-S}} & \textbf{\emph{NF-L}} & \textbf{\emph{F-S}} & \textbf{\emph{F-L}} \\
    \hline
    \hline
    \texttt{Encoder} & \multicolumn{4}{c|}{Emformer 68M} \\
    \hline
    \texttt{Predictor} & \multicolumn{4}{c|}{LSTM 6M} \\
    \hline
    $\texttt{Joiner}^b$ & \multirow{2}{*}{FC 5M} & \multirow{2}{*}{FC 11M} & FC 1K & FC 1M \\
    \cline{1-1} \cline{4-5}
    $\texttt{Joiner}^{nb}$ & & & FC 5M & FC 11M \\
    \hline
\end{tabular}
\\
\vspace{0.5em}
{\footnotesize\textbf{\textit{NF-S/L}}: non-factorized small/large}\space\space\space\space{\footnotesize\textbf{\textit{F-S/L}}: factorized small/large}\\
\caption{Component architecture and parameter breakdown of the RNN-T models used in this work. \emph{FC}: fully-connected layers.}
\label{table:model_params}
\end{table}

\def\JNO{{\textit{RTF}$_{\mathrm{join}}$}}
\def\RTFall{{\textit{RTF}$_{\mathrm{all}}$}}

\begin{table*}[t]
\centering
\begin{tabular}{ c c | c c c c c | c c c c c }
    \hspace*{-0.3em}{\texttt{thresh}}\hspace*{-0.3em} & $p^{\text{threshold}}$ & \textbf{\emph{test-clean}} & \textbf{\emph{test-other}} & \textbf{\emph{NBP}} & \hspace*{-0.3em}\textbf{\emph{\JNO}}\hspace*{-0.3em} & \hspace*{-0.3em}\textbf{\emph{\RTFall}}\hspace*{-0.3em} & \textbf{\emph{test-clean}} & \textbf{\emph{test-other}} & \textbf{\emph{NBP}} & \textbf{\emph{\JNO}} & \textbf{\emph{\RTFall}} \\
    \hline
    \hline
    & & \multicolumn{5}{c|}{\textbf{Non-Factorized Joiner Small (NF-S)}} & \multicolumn{5}{c}{\textbf{Non-Factorized Joiner Large (NF-L)}}\\
    \hline
     \multicolumn{2}{c|}{N/A} & 3.23 & 7.78 & 100\% & 0.11 & 0.31 & 2.97 & 7.33 & 100\% & 0.33 & 0.43 \\
    \hline
    \hline
    \multicolumn{2}{c|}{} & \multicolumn{5}{c|}{\textbf{Factorized Joiner Small (F-S)}} & \multicolumn{5}{c}{\textbf{Factorized Joiner Large (F-L)}}\\
    \hline
     16  & \hspace*{-0.3em}0.9999999\hspace*{-0.3em} & 3.26 & 7.68 & 99\% & 0.15 & 0.31 & 3.11 & 7.30 & 100\% & 0.33 & 0.42 \\
    %\hline
     8   & 0.9997 & 3.27 & 7.69 & 52\% & 0.08 & 0.25 & 3.10 & 7.31 & 50\% & 0.23 & 0.32 \\
    %\hline
     4   & 0.98 & 3.26 & 7.68 & 41\% & 0.07 & 0.23 & 3.09 & 7.30 & 41\% & 0.20 & 0.31 \\
    %\hline
     \textbf{2}   & \textbf{0.88} & \textbf{3.26} & \textbf{7.68} & \textbf{36\%} & \textbf{0.06} & \textbf{0.23} & \textbf{3.08} & \textbf{7.31} & \textbf{37\%} & \textbf{0.19} & \textbf{0.30} \\
    %\hline
     1   & 0.73 & 3.29 & 7.69 & 34\% & 0.06 & 0.23 & 3.12 & 7.36 & 35\% & 0.19 & 0.29 \\
    %\hline
     0.5 & 0.62 & 3.29 & 7.72 & 32\% & 0.06 & 0.23 & 3.13 & 7.41 & 34\% & 0.18 & 0.29 \\
    %\hline
     0.1 & 0.53 & 3.34 & 7.80 & 31\% & 0.06 & 0.22 & 3.13 & 7.50 & 33\% & 0.18 & 0.28 \\
\end{tabular}
\caption{Librispeech results. \textit{NBP}: non-blank percentage; \JNO: real-time factor of joiner only; \RTFall: total real-time factor.}
\label{table:libri_results}
\end{table*}

We conduct the majority of our experiments on the well-known Librispeech \cite{panayotov2015librispeech} corpus with 960 hours of labeled audio training data ($\sim$3K hours after speed perturbation~\cite{Ko2015AudioAF}). Accuracy evaluation (WER) is done on the \texttt{test-clean} and \texttt{test-other} splits. We also measure the model's runtime performance on 200 utterances randomly sampled from the two test splits with an actual Android device. The model is allocated two parallel execution threads, one for running the encoder, and one for running the predictor and joiner. We track the following runtime metrics:
\begin{itemize}
    \item \textbf{\emph{NBP}} (non-blank percentage): ratio of non-blank to blank joiner calls (in percentage). Note that for standard non-factorized joiners, this metric is always 100\%.
    \item \textbf{\emph{\JNO}} (joiner real time factor): the execution time of the joiner divided by the audio duration.
    \item \textbf{\emph{\RTFall}} (real time factor): the overall decoding time divided by the audio duration, i.e., the average decoding speed.
\end{itemize}

\subsubsection{Large-Scale In-House Data}
\label{sssec:in_house}

To better understand the effect of blank thresholding on WER, we also conduct experiments on our large-scale in-house dataset, which consists of 83M hand-transcribed de-identified utterances (145K hours) for training. We consider two evaluation sets:
\begin{itemize}
    \item \texttt{VA-P}: 14.9K hand-transcribed voice assistant utterances, collected via Meta Portal devices by internal volunteer participants who have agreed to having their Portal voice activity reviewed and analyzed.
    \item \texttt{VA-Q}: 44.2K hand-transcribed voice assistant utterances, collected via Meta Quest devices by a third party data vendor.
\end{itemize}

\subsection{Models and Training Procedure}

The focus of this paper is on the joiner, thus we keep the encoder, predictor, and output vocabulary constant across all RNN-T models. The encoder is a 20-layer streamable medium-latency Emformer~\cite{ShiWangWuEtAl21} with a stride of four, 240 ms lookahead, 1.28-second segments, input dimension 512, hidden dimension 2048, eight self-attention heads, and 1024-dimensional FC projection, a total of 68M parameters. The predictor consists of three LSTM layers with 512-dim hidden size, followed by 1024-dim FC projection, totaling 6M parameters. The output vocabulary is 5000 unigram SentencePieces~\cite{Kudo2018SubWord} estimated from the training transcripts, plus the blank symbol.

We consider two different types of joiners, non-factorized (\textbf{\emph{NF}}) and factorized (\textbf{\emph{F}}), as described in Section~\ref{ssec:model_arch}. For each type, we experiment with a small (\textbf{\emph{S}}) and large (\textbf{\emph{L}}) version with different number of parameters (5M vs.~11-12M) to better understand the model's performance at various joiner sizes. The small version consists of a single FC projection layer into the output vocabulary. The large version adds six 1024-dim FC hidden layers for the non-factorized (\texttt{Joiner}) and factorized non-blank joiner ($\texttt{Joiner}^{nb}$), and one 1024-dim FC hidden layer for the factorized blank joiner ($\texttt{Joiner}^b$), before the final FC output projection. Following~\cite{VarianiRybachAllauzenEtAl20}, we use rectified linear unit (ReLU) activation for the non-factorized joiners and hyperbolic tangent (Tanh) activation for the factorized joiners. Table~\ref{table:model_params} summarizes the component architectures and parameter breakdown of the RNN-T models considered in this work.

We train all models using Alignment Restricted RNN-T loss (left buffer 0, right buffer 15)~\cite{Mahadeokar2021AR-RNNT}, where the alignment is provided by a chenone hybrid acoustic model (AM) \cite{Le2019Kulfi}; SpecAugment LD policy~\cite{ParkChanZhangEtAl19} is applied on-the-fly during training. The models are trained with 32 A100 GPUs for 120 epochs (Librispeech data) or 15 epochs (large-scale in-house data). Librispeech models are evaluated without any external LM, while in-house experiments utilize finite state transducer (FST) biasing and neural LM shallow fusion~\cite{Le2021deepshallow,le21_interspeech}.

In this paper, we parameterize $p^{\text{threshold}}$ using a logistic sigmoid function $\sigma(\texttt{thresh})$.
For factorized joiners, we vary \texttt{thresh} $\in$ \{16, 8, 4, 2, 1, 0.5, 0.1\}, corresponding to $p^{\text{threshold}}$ $\in$ \{0.9999999, 0.9997, 0.98, 0.88, 0.73, 0.62, 0.53\}, respectively. We use a beam size of 10 and all models are 8-bit (INT8) quantized before decoding.

\section{Results and Discussion}
\label{sec:results}

\begin{figure*}
\centering
\begin{subfigure}{\columnwidth}
  \centering
  \includegraphics[width=\columnwidth]{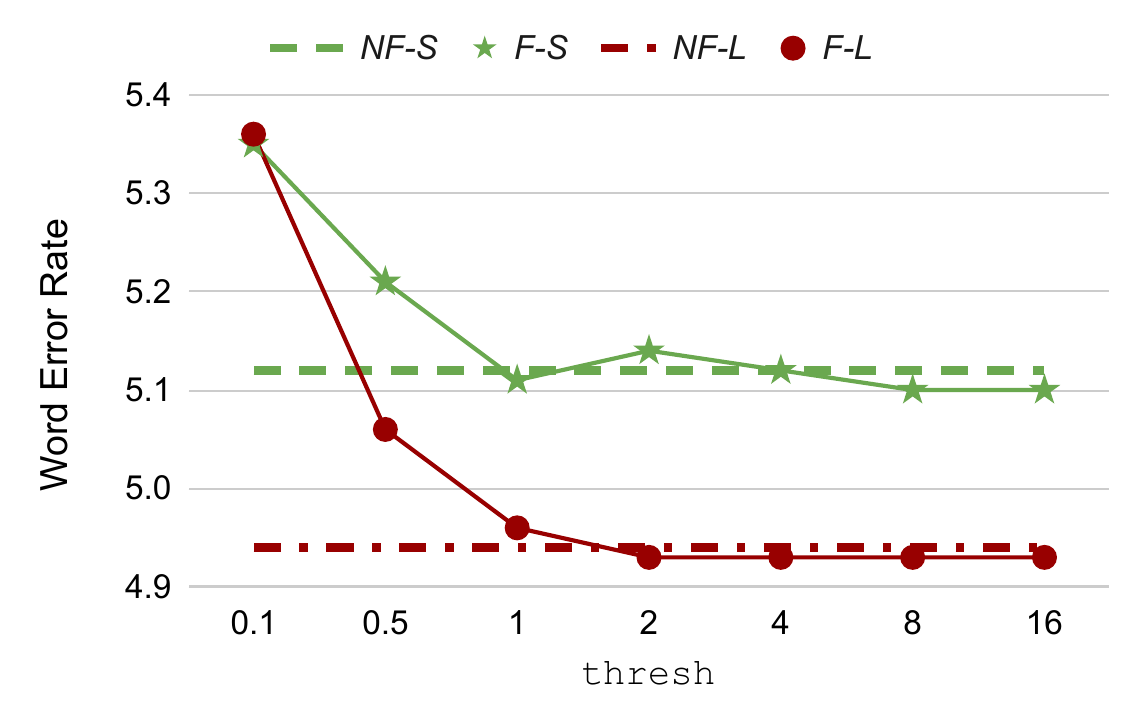}
  \caption{VA-P results}
  \label{fig:va1}
\end{subfigure}
\begin{subfigure}{\columnwidth}
  \centering
  \includegraphics[width=\columnwidth]{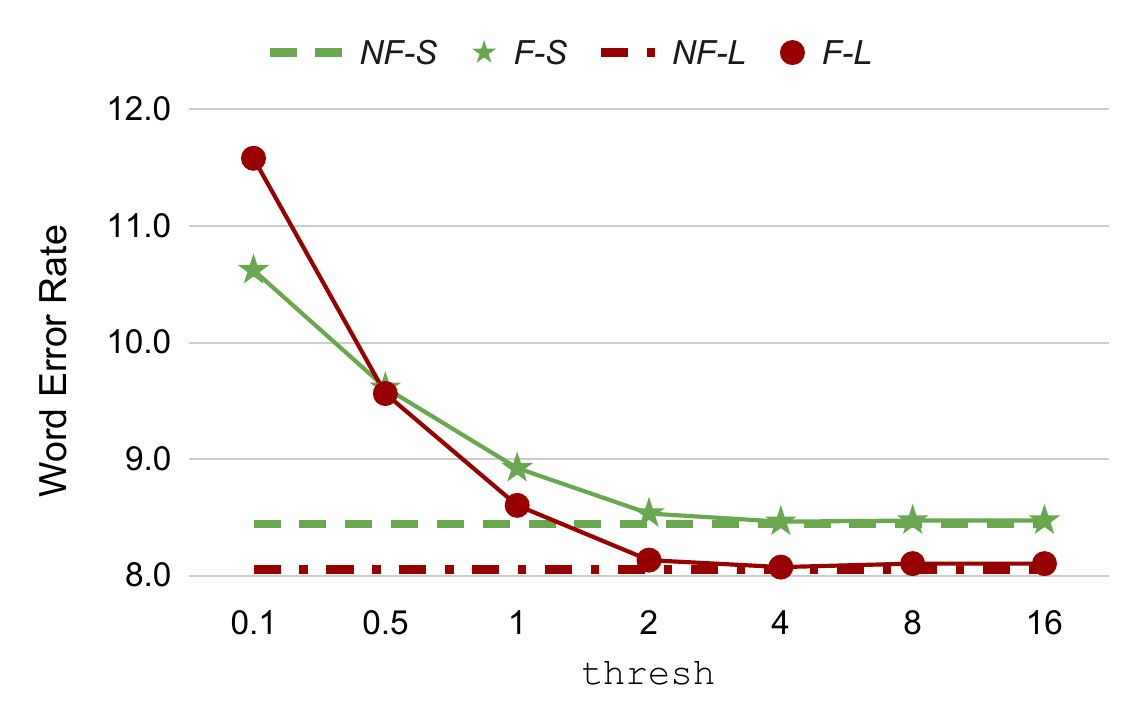}
  \caption{VA-Q results}
  \label{fig:va2}
\end{subfigure}
\caption{WER comparison between non-factorized and factorized joiners on in-house evaluation sets.}
\label{fig:in_house_results}
\end{figure*}

\subsection{WER and Runtime Efficiency}
\label{ssec:wer}

Table~\ref{table:libri_results} summarizes the results on Librispeech. Let us first examine the standard non-factorized joiners (\emph{NF-S} and \emph{NF-L}). Although the joiner in \emph{NF-S} makes up only 5M parameters and consists of a single FC layer, its overhead in decoding is non-trivial. This is because the joiner is invoked very often during decoding, roughly \textbf{80x} more than the encoder and \textbf{10x} more than the predictor. This explains why by going from \emph{NF-S} to \emph{NF-L}, which adds only 6M parameters to the joiner, \JNO~increases {\bf 3-fold} and \RTFall~grows by \textbf{39\%}. Note that enlarging the joiner leads to \textbf{8\%} and \textbf{6\%} relative WER reduction (WERR) on \texttt{test-clean} and \texttt{test-other}, respectively. This motivates why reducing \JNO~is important, as it not only improves \RTFall~but also allows us to utilize larger joiners and reduce WER.

Let us now focus on the proposed factorized joiners (\emph{F-S} and \emph{F-L}). The ``spikiness" of RNN-T's blank distribution is apparent through the NBP metrics at different \texttt{thresh} and $p^{\text{threshold}}$ values; roughly \textbf{50\%} of all blank probabilities are larger than 0.9997. Because of this spikiness, blank thresholding is effective at reducing the number of computationally expensive and unnecessary non-blank joiner calls. At \texttt{thresh} = 2, roughly \textbf{two-thirds} of all non-blank joiner invocations can be skipped, resulting in \textbf{42-45\%} reduction in \JNO~and \textbf{26-30\%} reduction in \RTFall~compared to the non-factorized baselines. Most crucially, blank thresholding has little to no impact on WER. Until \texttt{thresh} = 0.5, there is no visible WER degradation on either \texttt{test-clean} or \texttt{test-other} (defined as 1\% or higher relative WER increase) compared to \texttt{thresh} = 16, which effectively turns off blank thresholding. We will show in Section~\ref{ssec:in_house_results} that this robustness against WER degradation also extends to our large-scale in-house models and is not specific to Librispeech.

In summary, Table~\ref{table:libri_results} clearly demonstrates that the factorized joiner, when combined with blank thresholding, offers significantly better WER/RTF tradeoff compared to the standard non-factorized version. The former can achieve similar WER at \textbf{26-30\%} better RTF (\texttt{thresh} = 2, \emph{F-S} vs.~\emph{NF-S} and \emph{F-L} vs.~\emph{NF-L}). Viewed from a different angle, the former can achieve \textbf{6-8\%} better WER at similar RTF (\texttt{thresh} = 2, \emph{F-L} vs.~\emph{NF-S}). Our proposed method is also relatively simple to implement, involving only minor modifications to the joiner architecture and beam search algorithm.

\subsection{Large-Scale In-House Results}
\label{ssec:in_house_results}

Figure~\ref{fig:in_house_results} shows results on our two in-house evaluation sets, \texttt{VA-P} and \texttt{VA-Q}. Consistent with Librispeech, enlarging the non-factorized joiner leads to 4-5\% WERR across both evaluation sets (\emph{NF-L} vs.~\emph{NF-S}). The factorized joiner is also consistently on-par with the non-factorized baseline (\emph{F-S} vs.~\emph{NF-S} and \emph{F-L} vs.~\emph{NF-L}), demonstrating the former's efficacy on more realistic use cases. WER for factorized joiners is generally quite stable w.r.t.~\texttt{thresh}, although less than on Librispeech; this may be due to FST biasing and neural LM shallow fusion introducing additional score fluctuation.

As \texttt{thresh} decreases, WER on \texttt{VA-Q} degrades faster than on \texttt{VA-P}. A key difference of \texttt{VA-Q} compared to \texttt{VA-P} is that the former is considered out-of-domain w.r.t.~our training sets (acoustically and lexically), whereas the latter is in-domain. As a result, the model's confidence on \texttt{VA-Q} is likely lower and the score difference between competing hypotheses smaller. Thus, blank thresholding has more influence on the decoding results. This points to a limitation of our current approach, where the fixed threshold may not work well in all situations. We will address this limitation in future work.

\subsection{On-Device Power Consumption}
\label{ssec:power}

\begin{figure}[t]
  \centering
  \includegraphics[width=\columnwidth]{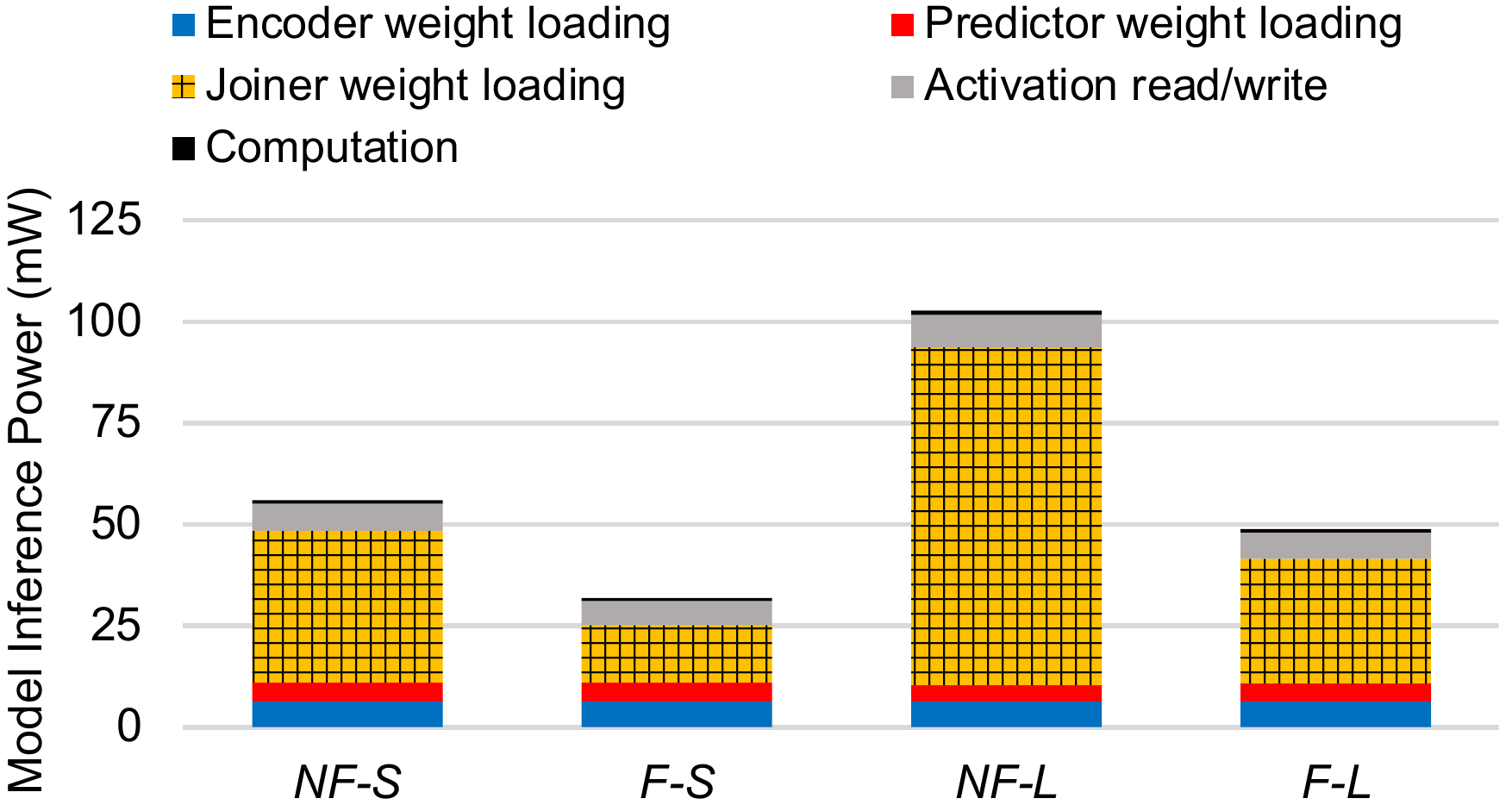}
  \caption{On-device power consumption breakdown.}
  \label{fig:power}
\end{figure}

We discovered that for on-device processing, the memory power for weight loading, 
 especially for loading the weights of the joiner, has dominated the total model inference power consumption, 
as shown in Figure~\ref{fig:power}. This can be attributed to two main reasons. 
First, the joiner is invoked much more frequently than the other RNN-T components, as mentioned in Section~\ref{ssec:wer}, since it needs to be run for every new encoder and predictor output. Second, assuming INT8 quantization, the 5MB \emph{NF-S} joiner and 11MB \emph{NF-L} joiner are too large to fit in energy-efficient local SRAM buffers (at most 1-2MB) of common mobile/edge DSP/accelerators, and loading weights from global DDR memory can be orders of magnitude less efficient. 

Our proposed factorized joiners (\emph{F-S} and \emph{F-L}) address both of the above inefficiencies, and subsequently, reduce the runtime power consumption by \textbf{43\%} and \textbf{53\%}, respectively, compared with their non-factorized counterparts (\emph{NF-S} and \emph{NF-L}). 
First, as most of the output tokens are blank, the lightweight factorized blank joiner can reduce the number of more costly non-blank joiner invocations by up to 69\%, as shown in Table~\ref{table:libri_results}. Second, we engineered the non-factorized blank joiners with suitable sizes (1KB for \emph{F-S} and 1MB for \emph{F-L}) to fit in the local memory of processors, which provides orders of magnitude power saving over global DDR memory.

The model inference power is estimated by collecting key runtime statistics (e.g., number of invocations for each model) and using hardware cost parameters of 120 pJ/byte access energy for DDR~\cite{li2019chip}, 1.5 pJ/byte access energy for local SRAM buffer~\cite{li2019chip}, and computation efficiency of 5 GOPS/mW (INT8)~\cite{lee2018unpu}. Note that though the computation efficiency can vary across different hardware backends (e.g., ASIC accelerator, DSP, and mobile CPU), the power reduction w.r.t.~total memory power should be consistent across different backends. Overall, our optimization can be critical to extend the operating time of battery-powered devices.

\section{Conclusion and Future Work}

In this work, we established the importance of RNN-T's joiner on transcription accuracy, decoding speed, and on-device memory consumption. To optimize the joiner, we utilize HAT-style joiner factorization and apply a novel blank thresholding technique to conditionally skip the more expensive non-blank computation. Our approach improves decoding speed by 26-30\% and reduces overall on-device power consumption by 43-53\%, all without any WER degradation. For future work, we plan to combine factorized blank thresholding with recent predictor optimization techniques to further improve runtime efficiency. We will also experiment with a dynamic and trainable probability threshold instead of using a fixed threshold.

\newpage
\footnotesize
\bibliographystyle{IEEEbib}
\bibliography{refs}

\begin{thebibliography}{10}

\bibitem{Graves12}
A.~Graves,
\newblock ``{Sequence Transduction with Recurrent Neural Networks},''
\newblock in {\em Proc. of ICML}, 2012.

\bibitem{GravesMohamedHinton13}
A.~Graves, A.~Mohamed, and G.~Hinton,
\newblock ``{Speech Recognition with Deep Recurrent Neural Networks},''
\newblock in {\em Proc. of ICASSP}, 2013.

\bibitem{RaoSakPrabhavalkar17}
K.~Rao, H.~Sak, and R.~Prabhavalkar,
\newblock ``{Exploring Architectures, Data and Units for Streaming End-to-End
  Speech Recognition with RNN-Transducer},''
\newblock in {\em Proc. of ASRU}, 2017.

\bibitem{YehMahadeokarKalgaonkar19}
C.~F. Yeh, J.~Mahadeokar, K.~Kalgaonkar, Y.~Wang, D.~Le, M.~Jain, K.~Schubert,
  C.~Fuegen, and M.~L. Seltzer,
\newblock ``{Transformer-Transducer: End-to-End Speech Recognition with
  Self-Attention},''
\newblock {\em arXiv preprint arXiv:1910.12977}, 2019.

\bibitem{ZhangLuSakEtAl20}
Q.~Zhang, H.~Lu, H.~Sak, A.~Tripathi, E.~McDermott, S.~Koo, and S.~Kumar,
\newblock ``{Transformer Transducer: A Streamable Speech Recognition Model with
  Transformer Encoders and RNN-T Loss},''
\newblock in {\em Proc. of ICASSP}, 2020.

\bibitem{ShiWangWuEtAl21}
Y.~Shi, Y.~Wang, C.~Wu, C.-F. Yeh, J.~Chan, F.~Zhang, D.~Le, and M.~L. Seltzer,
\newblock ``{Emformer: Efficient Memory Transformer Based Acoustic Model for
  Low Latency Streaming Speech Recognition},''
\newblock in {\em Proc. of ICASSP}, 2021.

\bibitem{Prabhavalkar17}
R.~Prabhavalkar, K.~Rao, T.~N. Sainath, B.~Li, L.~Johnson, and N.~Jaitly,
\newblock ``{A Comparison of Sequence-to-Sequence Models for Speech
  Recognition},''
\newblock in {\em Proc. of INTERSPEECH}, 2017.

\bibitem{HeSainathPrabhavalkarEtAl19}
Y.~He, T.~N. Sainath, R.~Prabhavalkar, I.~McGraw, R.~Alvarez, D.~Zhao,
  D.~Rybach, A.~Kannan, Y.~Wu, R.~Pang, Q.~Liang, D.~Bhatia, Y.~Shangguan,
  B.~Li, G.~Pundak, K.~C. Sim, T.~Bagby, S.-Y. Chang, K.~Rao, and A~Gruenstein,
\newblock ``{Streaming End-to-End Speech Recognition for Mobile Devices},''
\newblock in {\em Proc. of ICASSP}, 2019.

\bibitem{shangguan2019optimizing}
Y.~Shangguan, J.~Li, L.~Qiao, R.~Alvarez, and I.~McGraw,
\newblock ``{Optimizing Speech Recognition for the Edge},''
\newblock {\em arXiv preprint arXiv:1909.12408}, 2019.

\bibitem{Gulati2020conformer}
A.~Gulati, J.~Qin, C.~Chiu, N.~Parmar, Y.~Zhang, J.~Yu, W.~Han, S.~Wang,
  Z.~Zhang, Y.~Wu, and R.~Pang,
\newblock ``{Conformer: Convolution-Augmented Transformer for Speech
  Recognition},''
\newblock in {\em Proc. of INTERSPEECH}, 2020.

\bibitem{shi2022streaming}
Y.~Shi, C.~Wu, D.~Wang, A.~Xiao, J.~Mahadeokar, X.~Zhang, C.~Liu, K.~Li,
  Y.~Shangguan, V.~Nagaraja, et~al.,
\newblock ``{Streaming Transformer Transducer Based Speech Recognition Using
  Non-Causal Convolution},''
\newblock in {\em Proc. of ICASSP}, 2022.

\bibitem{Ghodsi20stateless}
M.~Ghodsi, X.~Liu, J.~Apfel, R.~Cabrera, and E.~Weinstein,
\newblock ``{RNN-Transducer with Stateless Prediction Network},''
\newblock in {\em Proc. of ICASSP}, 2020.

\bibitem{Rohit21LessIsMore}
R.~Prabhavalkar, Y.~He, D.~Rybach, S.~Campbell, A.~Narayanan, T.~Strohman, and
  T.~N. Sainath,
\newblock ``{Less is More: Improved RNN-T Decoding Using Limited Label Context
  and Path Merging},''
\newblock in {\em Proc. of ICASSP}, 2021.

\bibitem{Botros21TiedReduced}
R.~Botros, T.~N. Sainath, R.~David, E.~Guzman, W.~Li, and Y.~He,
\newblock ``{Tied \& Reduced RNN-T Decoder},''
\newblock in {\em Proc. of INTERSPEECH}, 2021.

\bibitem{VarianiRybachAllauzenEtAl20}
E.~Variani, D.~Rybach, C.~Allauzen, and M.~Riley,
\newblock ``{Hybrid Autoregressive Transducer (HAT)},''
\newblock in {\em Proc. of ICASSP}, 2020.

\bibitem{Sainath21Efficient}
T.~N. Sainath, Y.~He, A.~Narayanan, R.~Botros, R.~Pang, D.~Rybach, C.~Allauzen,
  E.~Variani, J.~Qin, Q.~Le-The, et~al.,
\newblock ``{An Efficient Streaming Non-Recurrent On-Device End-to-End Model
  with Improvements to Rare-Word Modeling},''
\newblock in {\em Proc. of INTERSPEECH}, 2021.

\bibitem{Chen22factorized}
X.~Chen, Z.~Meng, S.~Parthasarathy, and J.~Li,
\newblock ``{Factorized Neural Transducer for Efficient Language Model
  Adaptation},''
\newblock in {\em Proc. of ICASSP}, 2022.

\bibitem{GravesFernandezGomezEtAl06}
A.~Graves, S.~Fern{\'a}ndez, F.~Gomez, and J.~Schmidhuber,
\newblock ``{Connectionist Temporal Classification: Labelling Unsegmented
  Sequence Data with Recurrent Neural Networks},''
\newblock in {\em Proc. of ICML}, 2006.

\bibitem{Jain19decoding}
M.~Jain, K.~Schubert, J.~Mahadeokar, C.F. Yeh, K.~Kalgaonkar, A.~Sriram,
  C.~Fuegen, and M.~L. Seltzer,
\newblock ``{RNN-T for Latency Controlled ASR With Improved Beam Search},''
\newblock {\em arXiv preprint arXiv:1911.01629}, 2019.

\bibitem{panayotov2015librispeech}
V.~Panayotov, G.~Chen, D.~Povey, and S.~Khudanpur,
\newblock ``{Librispeech: An ASR Corpus Based on Public Domain Audio Books},''
\newblock in {\em Proc. of ICASSP}, 2015.

\bibitem{Ko2015AudioAF}
T.~Ko, V.~Peddinti, D.~Povey, and S.~Khudanpur,
\newblock ``{Audio Augmentation for Speech Recognition},''
\newblock in {\em Proc. of INTERSPEECH}, 2015.

\bibitem{Kudo2018SubWord}
T.~Kudo,
\newblock ``{Subword Regularization: Improving Neural Network Translation
  Models with Multiple Subword Candidates},''
\newblock in {\em Proc. of ACL}, 2018.

\bibitem{Mahadeokar2021AR-RNNT}
J.~Mahadeokar, Y.~Shangguan, D.~Le, G.~Keren, H.~Su, T.~Le, C.~Yeh, C.~Fuegen,
  and M.~L. Seltzer,
\newblock ``{Alignment Restricted Streaming Recurrent Neural Network
  Transducer},''
\newblock in {\em Proc. of SLT}, 2021.

\bibitem{Le2019Kulfi}
D.~Le, X.~Zhang, W.~Zheng, C.~Fuegen, G.~Zweig, and M.~L. Seltzer,
\newblock ``{From Senones to Chenones: Tied Context-Dependent Graphemes for
  Hybrid Speech Recognition},''
\newblock in {\em Proc. of ASRU}, 2019.

\bibitem{ParkChanZhangEtAl19}
D.~S. Park, W.~Chan, Y.~Zhang, C.-C. Chiu, B.~Zoph, E.~D. Cubuk, and Q.~V. Le,
\newblock ``{SpecAugment: A Simple Data Augmentation Method for Automatic
  Speech Recognition},''
\newblock in {\em Proc. of INTERSPEECH}, 2019.

\bibitem{Le2021deepshallow}
D.~Le, G.~Keren, J.~Chan, J.~Mahadeokar, C.~Fuegen, and M.~L. Seltzer,
\newblock ``{Deep Shallow Fusion for RNN-T Personalization},''
\newblock in {\em Proc. of SLT}, 2021.

\bibitem{le21_interspeech}
D.~Le, M.~Jain, G.~Keren, S.~Kim, Y.~Shi, J.~Mahadeokar, J.~Chan, Y.~Shangguan,
  C.~Fuegen, O.~Kalinli, Y.~Saraf, and M.~L. Seltzer,
\newblock ``{Contextualized Streaming End-to-End Speech Recognition with
  Trie-Based Deep Biasing and Shallow Fusion},''
\newblock in {\em Proc. of INTERSPEECH}, 2021.

\bibitem{li2019chip}
H.~Li, M.~Bhargava, P.~N. Whatmough, and H.-S.~P. Wong,
\newblock ``{On-Chip Memory Technology Design Space Explorations for Mobile
  Deep Neural Network Accelerators},''
\newblock in {\em Proc. of DAC}, 2019.

\bibitem{lee2018unpu}
J.~Lee, C.~Kim, S.~Kang, D.~Shin, S.~Kim, and H.-J. Yoo,
\newblock ``{UNPU: A 50.6 TOPS/W Unified Deep Neural Network Accelerator with
  1b-to-16b Fully-Variable Weight Bit-Precision},''
\newblock in {\em Proc. of ISSCC}, 2018.

\end{thebibliography}

\end{document}